\documentclass[aps,prb,reprint,preprintnumbers,showpacs,showkeys,superscriptaddress]{revtex4-1}
\usepackage{amsmath,amssymb,bm,mathrsfs,graphicx,longtable}
\usepackage[colorlinks=true,citecolor=blue,linkcolor=blue]{hyperref}

\begin{document}
\title{Electric Field-Induced Skyrmion Crystals via Charged Monopoles in Insulating Helimagets}

\author{Haruki Watanabe}
\email{hwatanabe@berkeley.edu}
\affiliation{Department of Physics, University of California, Berkeley, California 94720, USA}

\author{Ashvin Vishwanath}
\affiliation{Department of Physics, University of California, Berkeley, CA 94720, USA}
\affiliation{Materials Science Division, Lawrence Berkeley National Laboratories, Berkeley, CA 94720}

\begin{abstract}
Electrons propagating in a magnetically ordered medium experience an additional gauge field associated with the Berry phase of their spin following the local magnetic texture. In contrast to the usual electromagnetic field, this gauge field admits monopole excitations, corresponding to hedgehog defects of the magnetic order. In an insulator, these hedgehogs carry a well-defined electric charge allowing for them to be controlled by electric fields. One particularly robust mechanism that contributes to the charge is the orbital magnetoelectric effect, captured by a $\theta$ angle, which leads to a charge of $e\theta/2\pi$ on hedgehogs. This is a direct consequence of the Witten effect for magnetic monopoles in a $\theta$ medium. A physical consequence is that external electric fields can induce skyrmion crystal phases in insulating helimagnets.
\end{abstract}

\maketitle
\section{Introduction}
In magnetic materials, electrons follow the background spin texture due to the Hund's coupling and thereby acquire a Berry phase that plays the role of an emergent magnetic field. In contrast to the true electromagnetic field, this emergent gauge field admits magnetic monopoles, which are simply hedgehog defects of  the magnetic order. An advantage of realizing these monopoles is that electrons see these defects exactly as they would a true magnetic monopole, in contrast to those in spin ice \cite{Sondhi}, which are monopoles in $H$ rather than $B$. Furthermore, hedgehog defects are relatively common in magnets.  For example, recently, attention has focused on metallic helimagnets like MnSi, where broken inversion symmetry leads to spiral magnetic order. In certain regimes, a multi spiral  skyrmion lattice configuration is stabilized \cite{Rossler, BinzPRL,FischerRosch,Muhlbauer,Ong,
Tokura2012MnSi}. These correspond to a ground state with finite background density of emergent magnetic flux. There, hedgehog defects are observed in the formation and the melting process of the skyrmion crystal~\cite{Milde}. 

In the presence of a $\theta$ term in the electromagnetic action, $\mathcal{L}_{\theta}=(\theta/2\pi)(e^2/2\pi\hbar)\bm{E}\cdot\bm{B}$, magnetic monopoles acquire a fractional electric charge which is known as the Witten effect~\cite{Witten}.  In the condensed matter context, it is known that, in the absence of inversion and time-reversal symmetry, 3D insulators are partially characterized by the $\theta$ angle, that quantifies the electric polarization $\bm{P}=-(\theta/2\pi)(e^2/2\pi\hbar)\bm{B}$ in response to an applied orbital magnetic field~\cite{EssinMooreVanderbilt,EssinTurnerMooreVanderbilt,QiHugesZhang,RosenbergFranz,AriYiRogerAshvin}.  A hypothetical monopole $\nabla\cdot \bm{B}=(2\pi\hbar/e)\delta^3(\bm{x})$ would then be associated with an electric charge $\nabla\cdot\bm{P}=-e(\theta/2\pi)\delta^3(\bm{x})$, which has been used in numerical computations to diagnose the $\theta$ angle \cite{RosenbergFranz}.  However, in the absence of a physical realization of magnetic monopoles, this has remained a gedanken experiment. 

In this paper, we discuss physically realizable consequences of the Witten effect that leads to electrically charged hedgehogs in insulating magnets. An electrical insulator is required to allow us to sharply define charge. An important consequence we demonstrate is that the skyrmion crystal phase can be expanded, or even induced, by application of \emph{electric} fields.  We restrict ourselves to materials which display a diagonal magnetoelectric coupling $\bm{E}\cdot\bm{B}$~\cite{Sinisa_isotropic}, involving the {\em orbital} effect of applied magnetic fields. The orbital nature of the coupling is important since the emergent gauge field of the magnetic texture only has an orbital effect, although there can be other direct coupling of the magnetic texture to electrons without going through the emergent gauge field.  We also do not consider a part of magnetoelectric coupling $\bm{E}\cdot\bm{M}$ to single out the new effect. A more comprehensive theory is left for future work. Finally, we note that recently skyrmion lattices have been observed in the insulating magnet ${\mathrm{Cu}}_{2}{\mathrm{OSeO}}_{3}$~\cite{Seki1}. Although the $\theta$ angle is likely to be small in this particular material, in other systems where larger $\theta$  magneto electric angle may be realized, our work predicts that electric fields can facilitate the formation of skyrmion lattices and help  control magnetic properties in unusual ways.

\section{The $\theta$ term including the Berry phase}
Let us imagine the motion of electrons over a background spin texture $\bm{n}(\bm{x},t)$.  If the Hund's rule coupling $-J_H\bm{s}\cdot\bm{n}$ is much stronger than any other energy scales, the electron spin $\bm{s}$ will be slaved locally and instantaneously to $\bm{n}$.  In other words, as long as we are interested in the low-energy physics, we can project the electron spin onto the direction of $\bm{n}$.  As a consequence, electrons loose the spin degrees of freedom but instead experience the Berry vector potential $a_\mu=(\hbar/2e)(1-\cos\theta)\partial_\mu\phi$ [$(\theta,\phi)$ is the spherical coordinate of $\bm{n}$], in addition to the real U(1) gauge field $A_\mu$.  This means that, after integrating out these electrons, the resulting $\theta$ term takes the form
\begin{eqnarray}
\mathcal{L}_{\theta}&=&\frac{e}{\Phi_0}\frac{\theta}{2\pi}(\bm{E}+\bm{e})\cdot(\bm{B}+\bm{b}).
\label{thetaterm2}
\end{eqnarray}
Here, $\bm{E}$ and $\bm{B}$ are associated with the real U(1) gauge field $A_\mu$, while $\bm{e}$ and $\bm{b}$ are defined by the Berry phase $a_\mu$.  

\begin{figure}[t]
\begin{center}
\includegraphics[width=0.8\columnwidth,clip]{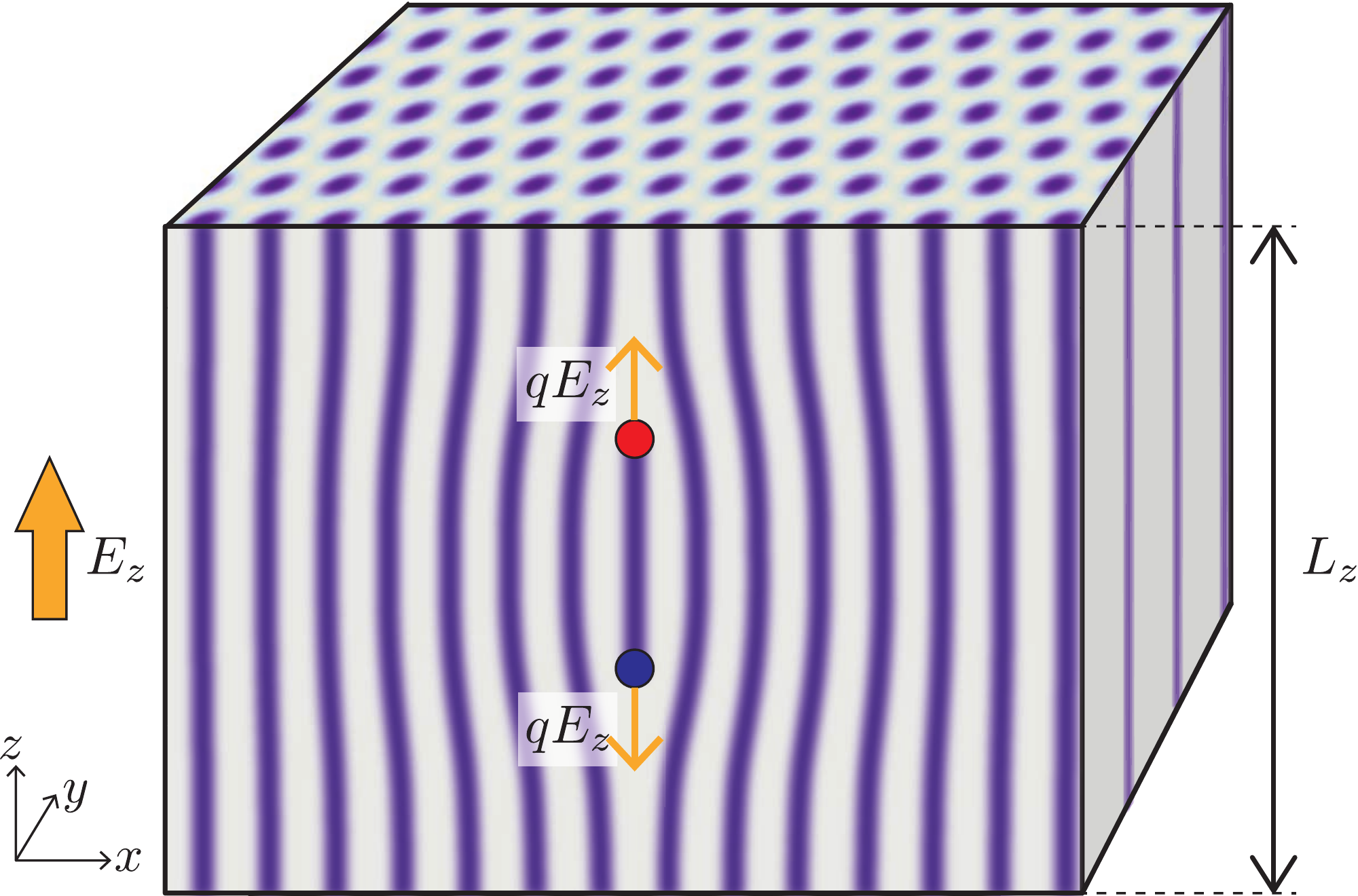}
\end{center}
\caption{(Color online) Illustration of the columnar-type 3D skyrmion crystal.  The blue lines represent skyrmions.  A hedgehog (red circle) and an anti-hedgehog (blue circle) appear at the two ends of a skyrmion line.  They feel the electric force $q\mathcal{E}_z$ in the opposite direction indicated by yellow arrows.}
\label{fig:nucleation}
\end{figure}

The $\theta$ term in Eq.~\eqref{thetaterm2} has the following important consequences. (i) A Hedgehog defect of the spin texture $\bm{n}$ acquires the electric charge
\begin{equation}
q=e\frac{\theta}{2\pi}Q[\bm{n}]
\label{hedgehogcharge}
\end{equation}
through the Witten effect, where $Q[\bm{n}]$ is the winding number of the defect defined by $Q[\bm{n}]\equiv(1/8\pi)\int_{S^2}\mathrm{d}S_i\epsilon^{ijk}\bm{n}\cdot\partial_j\bm{n}\times\partial_k\bm{n}\in\mathbb{Z}$.  Because of the  relation $b^i=-(\Phi_0/8\pi)\epsilon^{ijk}\bm{n}\cdot\partial_j\bm{n}\times\partial_k\bm{n}$, the hedgehog defect corresponds to a magnetic monopole of $\bm{b}$ with the magnetic charge $g=-\Phi_0Q[\bm{n}]$.  Therefore, from the Witten effect, we expect the electric charge $q=-e(\theta/2\pi)(g/\Phi_0)$.  Indeed, if integrated by parts, the cross term $\bm{E}\cdot\bm{b}$ of the $\theta$ term produces a term $e(\theta/2\pi)(\bm{\nabla}\cdot\bm{b}/\Phi_0)A_t$, which gives the electric charge in Eq.~\eqref{hedgehogcharge} from $q\equiv-\int\mathrm{d}^3x\partial\mathcal{L}_{\theta}/\partial A_t$.

(ii) The electric field $\bm{E}$ induces skyrmions in the plane orthogonal to $\bm{E}$.  To see this, let us compute the Hamiltonian associated with the $\theta$ term~\eqref{thetaterm2}:
\begin{eqnarray}
H_\theta&\equiv&\int\mathrm{d}^3x\left(\frac{\partial\mathcal{L_\theta}}{\partial\dot{\theta}}\dot{\theta}+\frac{\partial\mathcal{L_\theta}}{\partial\dot{\phi}}\dot{\phi}-\mathcal{L_\theta}\right)\notag\\
&=&-\frac{e}{\Phi_0}\frac{\theta}{2\pi}\int\mathrm{d}^3x\bm{E}\cdot(\bm{b}+\bm{B}).
\end{eqnarray}
In particular, if we set $\bm{E}=E_z\hat{\bm{z}}$ and $\bm{B}=\bm{0}$, $H_\theta$ reduces to a term of the form $-\mu_z N_z$, where
\begin{eqnarray}
\mu_z&\equiv&-\frac{\theta}{2\pi}eL_zE_z,\\
N_z&\equiv&\int\mathrm{d}^2x \frac{1}{4\pi}\bm{n}\cdot\partial_x\bm{n}\times\partial_y\bm{n},
\end{eqnarray}
where $N_z$ is the total number of skyrmions in each $xy$ plane and $L_z$ is the thickness of the system.  Here we have in mind a columnar-type skyrmion crystal as illustrated in Fig.~\ref{fig:nucleation}.  Therefore, $E_z$ acts as a chemical potential that induces or reduces skyrmions in $xy$ planes depending on its sign.

This effect can also be understood from the electric charge of hedgehogs explained above.  Suppose that a pair of a hedgehog ($Q[\bm{n}]=+1$) and an anti-hedgehog ($Q[\bm{n}]=-1$) is nucleated in the bulk.  The number of skyrmions is increased by one between the two defects (see Fig.~\ref{fig:nucleation})~\cite{Milde}. Since the hedgehog and the anti-hedgehog feel the electric force $qE_z$ in the opposite direction, if one applies a strong enough field in the right direction, they will be pulled apart and reach the opposite surfaces.  This way, the external electric field increases the skyrmion number $N_z$.

\begin{figure}[t]
\begin{center}
\includegraphics[width=\columnwidth,clip]{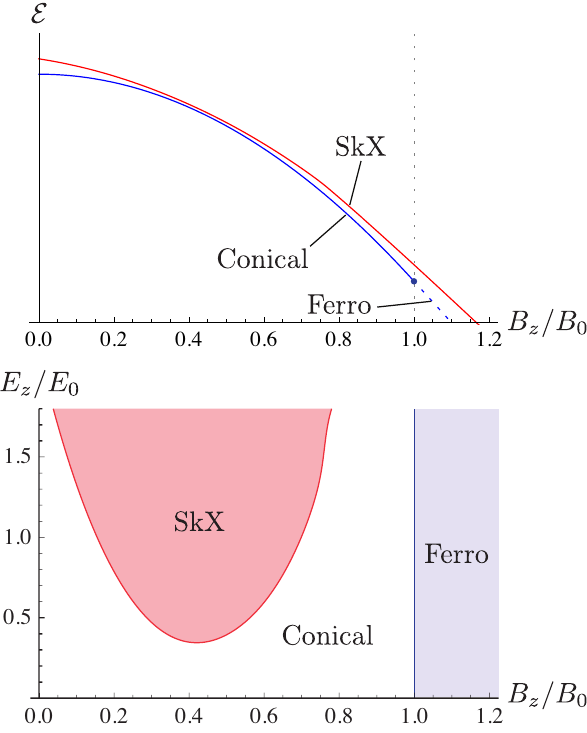}
\end{center}
\caption{(Color online) (a) The energy of the skyrmion crystal phase (red), the conical spin phase (blue), and the ferromagnet phase (dotted blue) as a function of the external magnetic field $B_z$ in the absence of the electric field.   (b) The phase diagram with the electric field $E_z$ and the magnetic field $B_z$.  The skyrmion crystal phase is induced by the external electric field as a consequence of the $\theta$ term.}
\label{fig:pd}
\end{figure}

To examine the effect of the external electric field in more detail, we take the simplest spin model
\begin{eqnarray}
\mathcal{E}&=&\int\mathrm{d}^3x\left[\frac{J}{2a}(\partial_i\bm{n})^2-\frac{D}{a^2}\bm{n}\cdot\bm{\nabla}\times\bm{n}-\mu B_z n_z\right]\notag\\
&&\quad-\mu_zN_z.\label{spinmodel}
\end{eqnarray}
Here, $J$ is the exchange constant and $D$ is the Dzyaloshinskii-Moriya coupling~\footnote{In this model, the skyrmion crystal phase tends to have a negative $N_z$}.  To avoid confusion with the electric field $\bm{E}$, here and hereafter we use $\mathcal{E}$ to indicate energies.   In Fig.~\ref{fig:pd} (a), we show the energy of the three competing magnetic orders, the skyrmion crystal phase (red), the conical spin phase (blue), and the ferromagnet phase (dotted blue), as a function of the external magnetic field $B_z/B_0$ [$B_0\equiv D^2/(\mu a^3J)$] in the absence of the electric field.  The energy of the conical spin phase is always slightly lower than that of the skyrmion crystal phase at $E_z=0$.  An external electric field of the order of $E_0\equiv 2\pi J/(ea_0\theta)$ can flip their order. Therefore, the skyrmion crystal phase may be induced as shown in Fig.~\ref{fig:pd} (b).

\section{Tight Binding Model with Tunable $\theta$}Thus far, we have treated the charge of the hedgehog as a phenomenological input parameter. Below, we explicitly demonstrate the fractional charge of hedgehogs in Eq.~\eqref{hedgehogcharge} using a simple tight-binding model of electrons propagating in a background magnetic order. We first discuss the band structure in the presence of ferromagnetic order, and then introduce the hedgehog defect. 

Let us first consider a fully polarized spin state and focus only on the single spin component of electrons.  To describe their band structure, we take 4 by 4 Hamiltonian~\cite{QiHugesZhang} in $3+1$ dimensions with a parameter $k_w\in[-\pi,\pi]$:
\begin{eqnarray}
&&H_{k_w}(\bm{k})=\sum_{\mu=0^4}\Gamma_\mu d_\mu,\label{toymodel}\\
\Gamma_\mu&=&(\tau_3,\tau_1\tilde{\sigma}_1,\tau_1\tilde{\sigma}_2,\tau_1\tilde{\sigma}_3,-\tau_2),\label{gamma}\\
d_\mu&=&(\sum_{i'}\cos k_{i'}-3,\sin k_x,\sin k_y,\sin k_z,\sin k_w).
\end{eqnarray}
In Eq.~\eqref{gamma}, $\tau,\tilde{\sigma}$'s are the Pauli matrices but they do not represent the electron spin; rather we look them as some internal degrees of freedom.  $\Gamma_\mu$'s satisfy the Clifford algebra $\{\Gamma_\mu,\Gamma_\nu\}=2\delta_{\mu\nu}\openone$ ($\mu,\nu\in\{0,\dots,4\}$).  We set the hopping constant $t=1$ and the lattice constant $a=1$ to simply equations. 

\begin{figure}[t]
\begin{center}
\includegraphics[width=\columnwidth,clip]{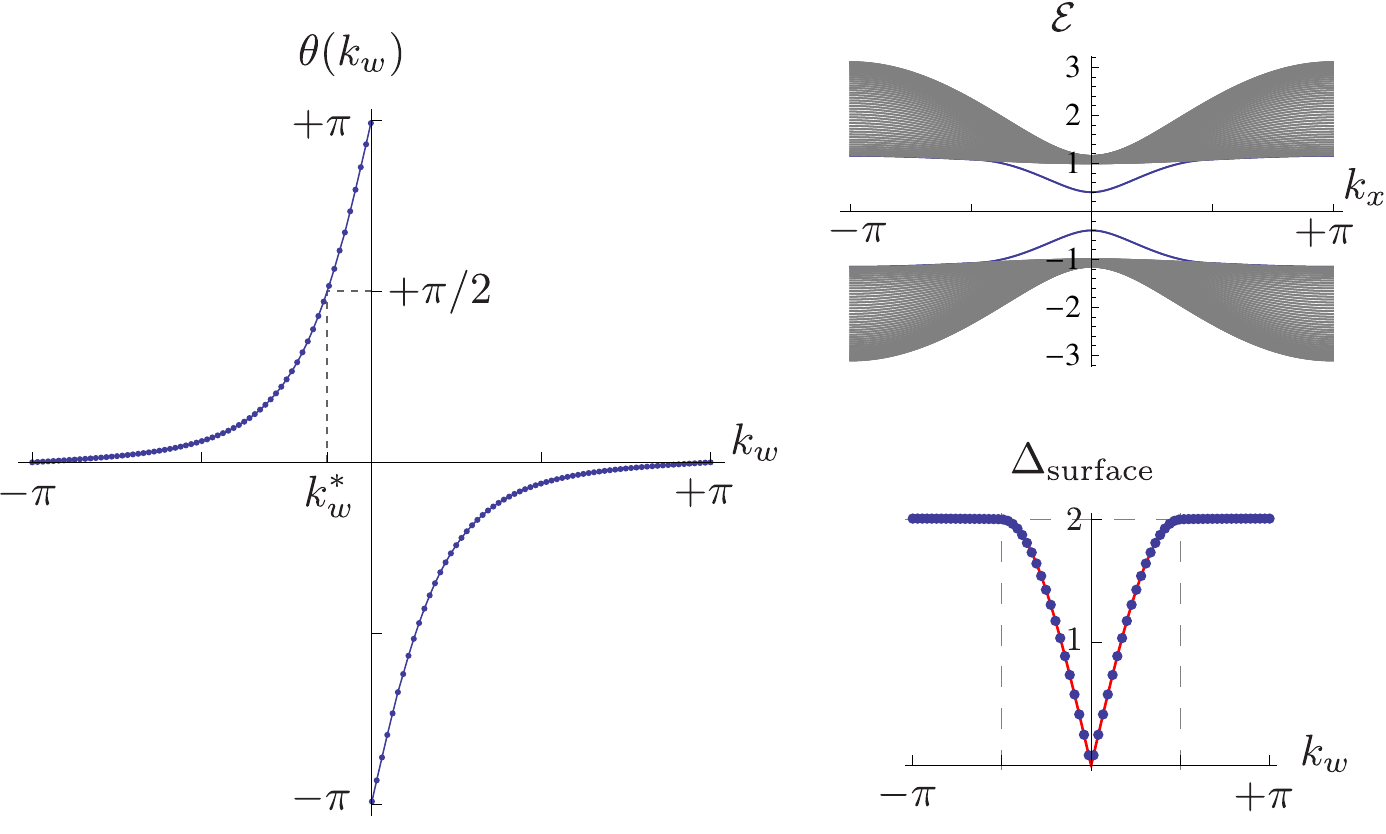}
\end{center}
\caption{(Color online) (a) The $\theta$ angle of the model \eqref{toymodel} as a function of the parameter $k_w$.  A the inversion symmetric points $k_w=0$ and $2\pi$, $\theta$ has to be quantized to either $0$ or $\pm\pi$.  (b) The spectrum of the same model with the open boundary condition in the $z$ direction.  For this plot we set $k_w=k_w^*$ and $k_y=0$.  There are gapped surface states inside the bulk gap $\Delta_{\text{bulk}}=2$. (c) The gap $\Delta_{\text{surface}}$ of the surface state as a function of $k_w$.  The solid red line represents $2|\sin k_w|$.}
\label{fig:thetaangle}
\end{figure}

The bulk spectrum  $\varepsilon(\bm{k})=\pm\sqrt{d_\mu d_\mu}$ has a gap $\Delta_{\text{bulk}}=2$ for any value of $k_w$ and we fill the doubly-degenerate bottom band.  The $\theta$ angle of this insulator can be obtained by evaluating the integral
\begin{equation}
\theta(k_w)=-\frac{1}{4\pi}\int\mathrm{d}^3k\epsilon^{ijk}\text{tr}\left(a_i\partial_ja_k+\frac{2i}{3}a_ia_ja_k\right),\label{thetaangle}
\end{equation}
where $(a_i)_{nm}\equiv-i\langle n k|\partial_{k_i}|mk\rangle$ ($n,m\in\{1,2\}$, $i,j,k\in\{1,2,3\}$).   We plot numerically computed $\theta(k_w)$ in Fig.~\ref{fig:thetaangle} (a).  As $k_w$ varies from $-\pi$ to $\pi$, $\theta(k_w)$ explores all values in $[-\pi,\pi]$. This is expected since one can view our model as a model in $4+1$ dimensions with a nonzero second Chern number $\nu_2=+1$~\cite{QiHugesZhang}.  $\theta$ is quantized to either $0$ or $\pm\pi$ at inversion symmetric points $k_w=0,\pi$.  With the open boundary condition, there is a gapped Weyl fermion on each surface with the gap $\Delta_{\text{surface}}=2|\sin k_w|$ (for $|k_w|\leq \pi/2$) [see Fig.~\ref{fig:thetaangle} (b) and (c)]. 

{\em Electronic Structure of Hedgehogs:} We now restore the electron spin.  Assuming the Hund's rule coupling to the background spin $\bm{n}_{\bm{x}}$, the total Hamiltonian in the real space reads
\begin{eqnarray}
[H_{\text{(tot)}}]_{\bm{x},\bm{x}'}&=&\left[(\cos k_w-3)\Gamma_0+\sin k_w\Gamma_4-\frac{J}{2}\bm{n}_{\bm{x}}\cdot\bm{\sigma}\right]\delta_{\bm{x},\bm{x}'}\notag\\
&&+\frac{\Gamma_0-i\Gamma_i}{2}\delta_{\bm{x}+\hat{\bm{i}},\bm{x}'}+\text{h.c.}\label{total}
\end{eqnarray}
Our model contains eight bands in total and we fill two lowest bands.  Note that the crystal symmetry of the cubic lattice allows only the diagonal couplings, like $\bm{E}\cdot\bm{b}$ and $\bm{E}\cdot\bm{B}$. Furthermore, the internal SU(2) symmetry of $H_{\text{(tot)}}$ prohibits the $\bm{E}\cdot\bm{n}$ coupling.

In the following, we consider three different background spin configurations: (i) the ferromagnetic order without defects $\bm{n}_{\bm{x}}=\hat{\bm{z}}$, (ii) with a hedgehog defect $\bm{n}_{\bm{x}}=(\bm{x}-\bm{x}_0)/|\bm{x}-\bm{x}_0|$, and (iii) with an anti-hedgehog defect $\bm{n}_{\bm{x}}=-(\bm{x}-\bm{x}_0)/|\bm{x}-\bm{x}_0|$.  We take a finite size system $L^3=10^3$ with the open boundary condition and $x,y,z\in\{-(L/2)+1,\ldots,L/2\}$ and we set $\bm{x}_0=(0.5,0.5,0.5)$ to avoid the singularity.  

\begin{figure}
\begin{center}
\includegraphics[width=\columnwidth,clip]{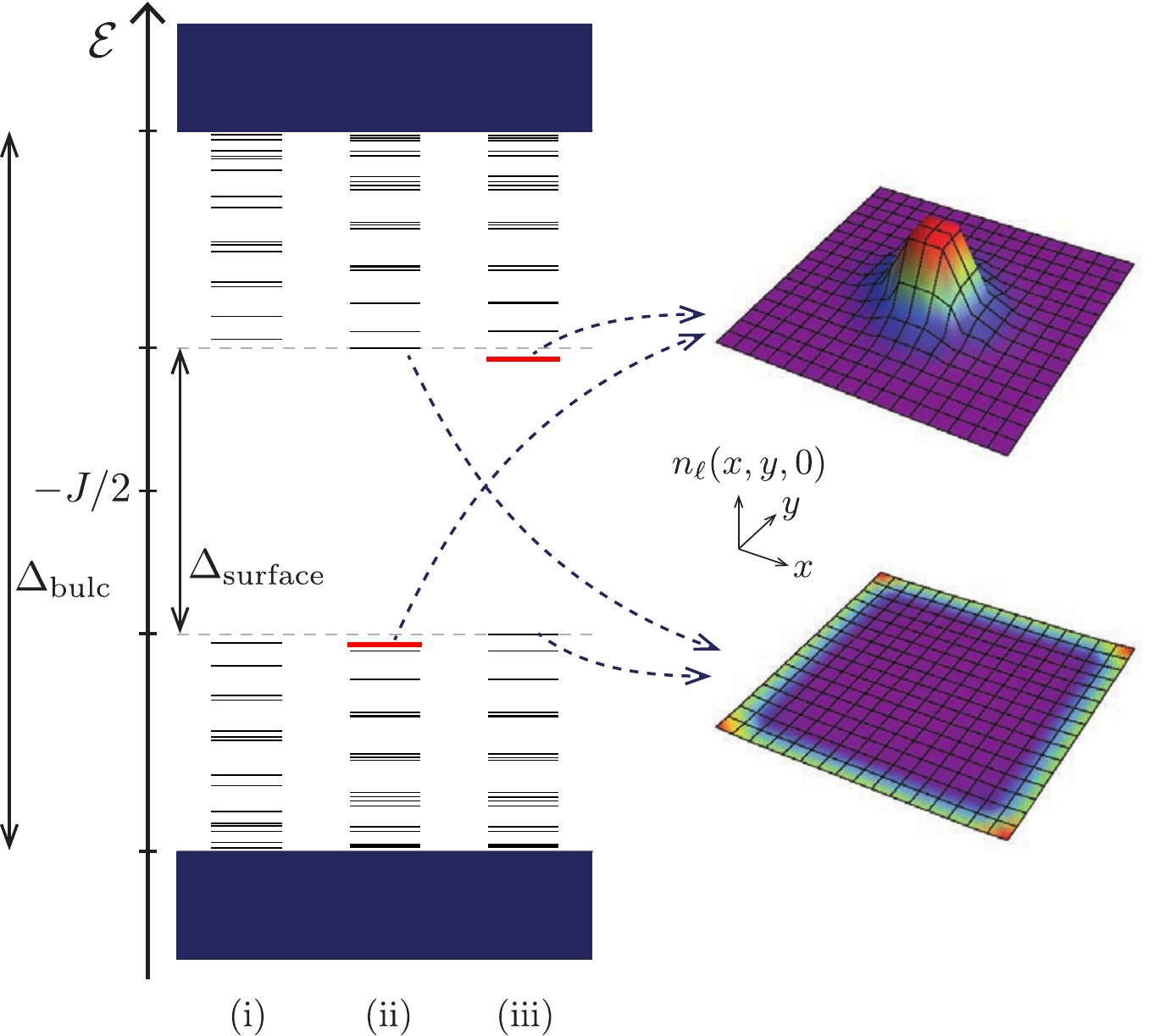}
\end{center}
\caption{(Color online) (a) The energy levels of the model \eqref{total} near the energy eigenvalue $\mathcal{E}= -J/2$ for each background spin texture (i), (ii), and (iii) (see the text).  The shaded region represents the bulk. Black lines are surface states (the discreteness is due to the finite system size), while the two red lines are those bound to the (anti-)hedgehog at the origin.  (b) The density profile of the bound states. We set $z=0$ for this plot. (c) For comparison we show the density profile of two surface states with a similar energy.}
\label{fig:levels}
\end{figure}

The left panel of Fig.~\ref{fig:levels} shows the energy levels near $E=-J/2$ for $J=50$ and $k_w=k_w^*\approx-0.409$ at which the $\theta$ angle is $+\pi/2$.  In the absence of defects [(i)], $J$ merely shifts the origin of the energy depending on the spin.  Even in the presence of a hedgehog [(ii)] or an anti-hedgehog [(iii)], the spectrum is not very different from that of the case (i) since the spin texture is still locally ferromagnetic. In particular, there remains the surface gap $\Delta_{\text{surface}}=2|\sin k_w^*|\approx 0.80$ even when the defect is present.  

The notable difference between (i) and (ii) [or (i) and (iii)] is the appearance of a state bound to the (anti-)hedgehog, which is denoted by the thick red lines in Fig.~\ref{fig:levels}.  The right panels of Fig.~\ref{fig:levels} illustrate the density profile of eigenstates near the surface gap.  Here, the density profile of the $\ell$-th eigenstate is defined by $n_\ell(\bm{x})\equiv\sum_{\alpha=1}^8|\psi_\ell(\bm{x},\alpha)|^2$ using the eigenfunction $\psi_\ell(\bm{x},\alpha)$ ($\ell\in \{1,2,\ldots,8L^3\}$, $\alpha\in \{1,2,\ldots,8\}$).
The density of the bound state is clearly peaked at the origin, while all other mid-gap states are localized at the surface.  The bound state appears below (above) the surface gap for the hedgehog (anti-hedgehog) configuration. 

\begin{figure}
\begin{center}
\includegraphics[width=\columnwidth,clip]{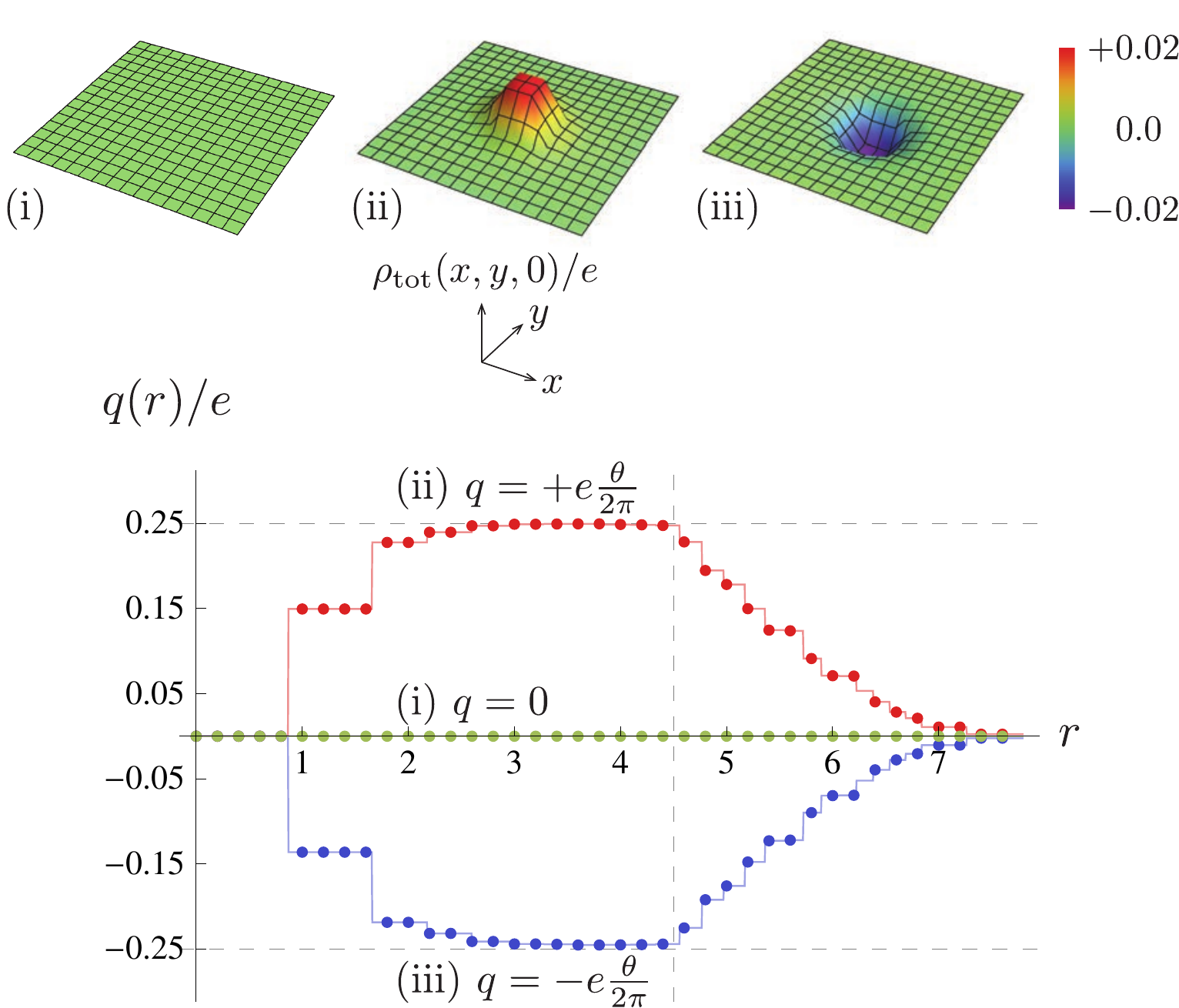}
\end{center}
\caption{(Color online) (a) The total charge density Eq.~\eqref{totalcharge} with or without (anti-)hedgehogs.  We set the chemical potential at $\mu=-J/2$ (see Fig.~\ref{fig:levels}).  There is a charge excess (deficiency) at the hedgehog (anti-hedgehog).  (b) The integral charge as a function of the distance from the origin [Eq.~\eqref{integralcharge}].  $q(r)$ first saturates at the expected value and then starts to decrease at $r= (L-1)/2$ due to the surface contribution.}
\label{fig:charge}
\end{figure}

Let us now set the chemical potential inside the surface gap and compute the net charge density of the resulting state:
\begin{eqnarray}
\rho_{\text{tot}}(\bm{x})&\equiv&e\sum_{\ell:\text{filled}}\left[n_\ell(\bm{x})-L^{-3}\right].\label{totalcharge}
\end{eqnarray}
The second term is the contribution of ions that neutralizes the total electric charge of the system.  Figure~\ref{fig:charge} (a) shows the profile of the total charge density for spin configuration (i), (ii), and (iii), respectively.  One can clearly see the charge excess (deficiency) at the hedgehog (anti-hedgehog).

Finally, Fig.~\ref{fig:charge} (b) shows the integral charge 
\begin{equation}
q(r)\equiv \sum_{\bm{x}}\vartheta(r-|\bm{x}-\bm{x}_0|)\rho_{\text{tot}}(\bm{x})\label{integralcharge}
\end{equation}
as a function of the distance from the defect, where $\vartheta$ is the step function.  Here, $k_w$ is still fixed to $k_w^*$ that corresponds to $\theta=\pi/2$. Clearly, $q(r)$ saturates at its expected value $q=+e/4$ for a hedgehog $Q[\bm{n}]=+1$ and $q=-e/4$ for an anti-hedgehog $Q[\bm{n}]=-1$ around $r=3$.  The integral charge then starts to decrease at $r= (L-1)/2=4.5$ due to the contribution from surface states.  This is expected since the net electric charge of the system has to be neutral.   One can avoid the contribution from the surface by taking the periodic boundary condition. To this end there must be the same number of hedgehogs and anti-hedgehogs, as in Fig.~\eqref{fig:nucleation}.

\begin{figure}
\begin{center}
\includegraphics[width=0.8\columnwidth,clip]{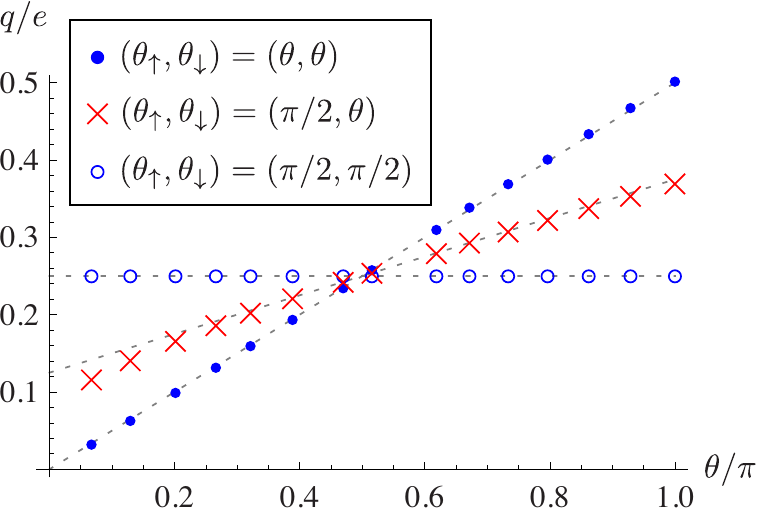}
\end{center}
\caption{(Color online) The electric charge $q$ of the hedgehog defect, defined by $\max_{r}q(r)$ in Eq.~\eqref{integralcharge}. When $\theta_{\uparrow}=\theta_{\downarrow}(=\theta)$, the electric charge agrees with $q/e=\theta/2\pi$ (dotted lines). As an example of the case $\theta_{\uparrow}\neq\theta_{\downarrow}$, the red crosses show the electric charge $q$ for $\theta_{\uparrow}=\pi/2$ and $\theta_{\downarrow}=\theta$. The charge is well fit by $q/e=(\theta_{\uparrow}+\theta_{\downarrow})/4\pi$ (the dotted line).}
\label{fig:charge12}
\end{figure}

Let us make three remarks in order.  (1) When one adds one more electron to the case (iii) and fills the bound state to the anti-hedgehog, the electric charge of the anti-hedgehog will become 0.75.  Similarly, if one subtracts one electron from the case (ii), the electric charge of the hedgehog will become $-0.75$. (2) The electric charge of hedgehogs may deviate from the idealized value in Eq.~\eqref{hedgehogcharge} for a smaller value of $J$, but we checked numerically that the charge varies only about 10 $\%$ even when $J$ is as small as $5$, which is the order of the band width. (3) So far we have considered the simplest case where the $\theta$ angles for the spin up and down electrons $\theta_{\uparrow,\downarrow}$ [before performing the local SU(2) rotation] are the same.  In general, however, in the presence of the Zeeman coupling and lattice anisotropy, they may take different values. In such a case, the effective $\theta$ angle should be given by the average of $\theta_{\uparrow}$ and $\theta_{\downarrow}$. Namely, the charge of the hedgehog is given by $q=(\theta_{\uparrow}+\theta_{\downarrow})/4\pi$.  We confirm this numerically in Fig.~\ref{fig:charge12}.

\section{Application to real materials}
Let us now discuss the experimental feasibility of above phenomena in real magnetic materials.  First of all, the minimum field required to induce the skyrmion crystal phase in Fig.~\ref{fig:pd} $E_z\simeq 0.33E_0=3$~kV$/$mm for $\theta=\pi$, $J=2$~meV, $a=4$~\AA.  This is strong but not uncommon in the real experiment.  In fact, an electric field of this order of strength has already been applied to the skyrmion lattice phase of the insulating helimagnet ${\mathrm{Cu}}_{2}{\mathrm{OSeO}}_{3}$~\cite{White, Omrani} to explore the megnetoelectric effect of the material~\cite{Maisuradze,Belesi,Seki1,Seki2}. The absence of a shift in the skyrmion crystal phase boundary in these electric fields indicates that the orbital  $\theta$ angle is much smaller than $O(1)$ in this particular material. However, ideas to realize larger $\theta$ eg. by proximity to topological states have been discussed \cite{Sinisa_isotropic,AriYiRogerAshvin}

Realizing a skyrmion crystal with a large $\theta$-angle in an actual experiment may require two independent mechanisms for each.  For example, one can follow the guiding principle proposed in Ref.~\onlinecite{Sinisa_isotropic} to get an $O(1)$ isotropic magnetoelectric coupling. In addition we would require an independent set of magnetic moments that would create the skyrmion texture.

The response of the magnetic texture $\bm{n}$ toward the applied electric field $\bm{E}$ does not solely come from the $\theta$ term in Eq.~\eqref{thetaterm2}. In fact, in magnetoelectric materials, $\mathcal{L}'=\chi\bm{E}\cdot\bm{M}$ ($\bm{M}=|\bm{M}|\bm{n}$) is a more familiar term. It contains $\chi(\bm{\nabla}\cdot\bm{M})A_t$ after an integration by parts, so that any inhomogeneous magnetic texture acquires an electric charge density $\rho=-\chi\bm{\nabla}\cdot\bm{M}$~\cite{Nomura}.  However, in an experiment on ${\mathrm{Cu}}_{2}{\mathrm{OSeO}}_{3}$~\cite{Omrani}, the magnetization is found to be order of $10^{-6}\mu_B/\text{Cu}^{2+}$ under $\sim 8$ kV$/$mm electric field.  This implies that $\chi$ is the order of $10^{-4}$ in the unit of $e^2/2\pi\hbar$ and that the effect of $\mathcal{L}'$ can be subdominant if $\theta$ is $O(1)$, depending on specific materials.

\section{Conclusion}
In this paper, we demonstrated that hedgehog defects in magnets can play the role of magnetic monopoles via the Hund's coupling to the electron's spin.  As a result, they acquire the electric charge $q=(\theta/2\pi)Q[\bm{n}]$ ($Q[\bm{n}]$ is the winding number) through the Witten effect.  This opens a way to control skyrmions using \emph{electric} fields: an electric field of the order $\sim 1$ kV$/$mm may induce a skyrmion crystal phase in 3D helimagnets.  The only essential ingredients behind these effects are the strong Hund's coupling and a fairly large $\theta$ angle that describes the diagonal orbital magnetoelectric response. This provides additional impetus to the search for magnetic materials with large orbital magneto electric coupling. 

\acknowledgements
We thank Oleg Tchernyshyov for useful comments.  This research was supported by the ARO MURI on topological insulators grant W911-NF-12-1-0961.
\bibliography{references}
\end{document}